\documentclass[10pt,a4paper]{article}

\usepackage{times}
\usepackage{epsfig,wrapfig}
\usepackage{epstopdf} 

\usepackage{fancyhdr}

\usepackage[lmargin=2.5cm, rmargin=2.5cm,tmargin=3.50cm,bmargin=3.50cm]{geometry}
\usepackage{indentfirst}

\usepackage{titlesec}
\titleformat{\section}[hang]
  {\centering}{\thesection}{1ex}{\normalsize \textsc}
\titleformat{\subsection}[hang]
  {}{\thesubsection}{1ex}{\normalsize \textit}

\renewcommand{\thesection}{ \normalsize \textnormal{\Roman{section}.}}
\renewcommand{\thesubsection}{\normalsize \textnormal{\textsc{\textit{\Alph{subsection}.}}}}

\def\e{\begin{equation}}
\def\f{\end{equation}}
\def\_#1{{\bf #1}}

\def\.{\cdot}

\usepackage{bm}
\usepackage[usenames,dvipsnames]{xcolor}
\usepackage{srcltx}
\usepackage{psfrag}
\usepackage{epsfig}
\usepackage{graphicx}
\usepackage{epstopdf}
\usepackage{color}
\usepackage[tbtags,sumlimits,nointlimits,reqno]{amsmath}
\usepackage{amssymb}
\usepackage{color}
\usepackage{float}
\usepackage{subfigure}
\usepackage{gensymb}
\usepackage[mode=errorstop]{pstool}
    \newcommand{\figref}{Fig.~\ref}

\begin{document}

\title{\large \textbf{Nonreciprocal Phase Gradient Metasurface: Principle and Transistor Implementation}}
%
\def\affil#1{\begin{itemize} \item[] #1 \end{itemize}}
\author{\normalsize \bfseries \underline{G. Lavigne}$^1$ and C. Caloz$^1$}
\date{}
\maketitle
\thispagestyle{fancy} 
\vspace{-6ex}
\affil{\begin{center}\normalsize $^1$Polytechnique Montr\'eal, Department of Electrical Engineering, Blvd. Edouard-Montpetit, H3T 1J4, Montr\'eal, Canada  \\
guillaume.lavigne@polymtl.ca
 \end{center}}

\begin{abstract}
\noindent \normalsize
\textbf{\textit{Abstract} \ \ -- \ \
We introduce the concept of nonreciprocal nongyrotropic phase gradient metasurfaces. Such metasurfaces are based on bianisotropic phase shifting unit cells, with the required nonreciprocal and nongyrotropic characteristics. Moreover, we present a transistor-based implementation of a nonreciprocal phase shifting subwavelength unit cell. Finally, we demonstrate the concept with a simulation of a 6-port spatial circulator application. 
}
\end{abstract}

\vspace{-5mm}
\section{Introduction}
\vspace{-2.5mm}
Metasurfaces have been shown to enable a myriad of wave transformations~\cite{achouri2018design,asadchy2018bianisotropic}. The vast majority of the works reported so far in this area have focused on reciprocal structures. Integrating of nonreciprocity~\cite{caloz2018electromagnetic} in metasurfaces inherently bears potential for even more sophisticated and useful metasurface wave transformations. A few nonreciprocal metasurfaces have already been recently reported, based either on time-varying~\cite{shaltout2015time,shi2017optical} or on transistor-loaded structures~\cite{kodera2011artificial,ra2016magnet,taravati2017nonreciprocal}.

We introduce here the concept of nonreciprocal nongyrotropic phase gradient metasurfaces based on nonreciprocal phase shifting unit cells. Moreover, we propose a transistor-based implementation of such a structure, and demonstrate it in a 6-port spatial circulator.

\vspace{-5mm}
\section{Operation Principle}
\begin{figure}[h]
  \centering
  \includegraphics[width=0.8\textwidth]{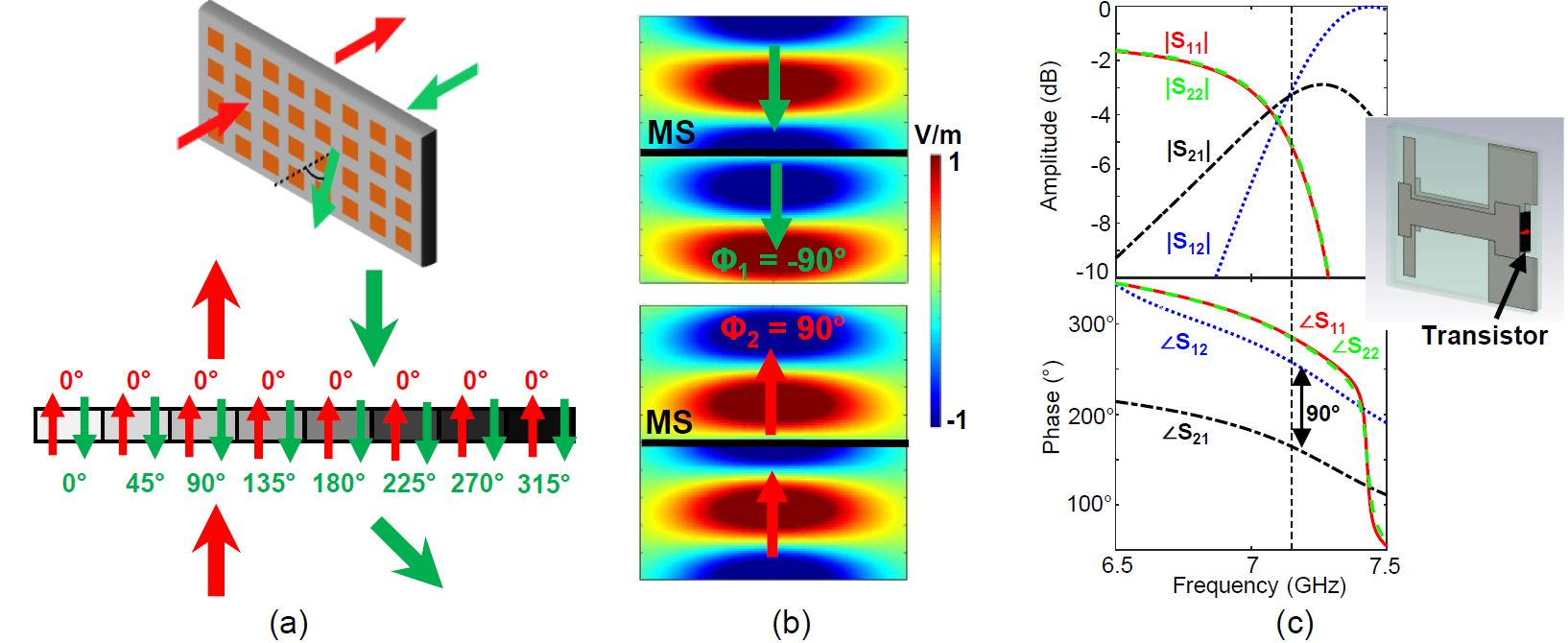}
  \caption{Nonreciprocal phase gradient metasurface based on nonreciprocal phase shifting unit cells. (a)~Illustration of a nonreciprocal phase gradient metasurface composed of 8 discretized nonreciprocal phase shifters. (b)~ FDFD-simulated real part of $E_x$ of a nonreciprocal phase shifting unit cell with $\phi_1=-90^\circ$ and $\phi_2 =+90^\circ$ ($|\Delta \phi| = 180^\circ$). (c)~Full-wave simulated scattering parameters (CST Microwave Studio) for a transistor-based implementation of a uniform nonreciprocal phase shifting metasurface with $|\Delta \phi| = 90^\circ$.}\label{fig:Concepts}
\end{figure}

\vspace{-2.5mm}
The proposed metasurface is based on nonreciprocal phase shifting particles arranged in such a way as to exhibit a phase gradient for wave deflection in space, as illustrated in~\figref{fig:Concepts}~(a). In the case of this figure, the phase gradient has been set only in one direction of propagation, so that the wave trajectory is deflected in that direction, while being unaltered in the opposite direction.
%

\vspace{-2.5mm}
\section{Nonreciprocal Phase Shifting Metasurface}
\vspace{-2.5mm}
For simplicity, we assume here nongyrotropy, and we will report on the gyrotropic case elsewhere. Except for this restriction, we allow the metasurface to be bianisotropic, which leads to the tangential susceptiblity-based generalized sheet transition conditions (GSTCs)~\cite{achouri2018design}

\begin{subequations}\label{eq:GSTC}
\begin{equation}
\hat{z} \times \Delta\mathbf{H} = j \omega \epsilon \overline{\overline{ \chi}}_\text{ee} \mathbf{E}_\text{av} +  j k \overline{\overline{ \chi}}_\text{em}   \mathbf{H}_\text{av} ,
\end{equation}
\begin{equation}
\Delta \mathbf{E} \times \hat{z}   = j k \overline{\overline{ \chi}}_\text{me}  \mathbf{E}_\text{av} + j \omega \mu \overline{\overline{ \chi}}_\text{mm} \mathbf{H}_\text{av},
\end{equation}
\end{subequations}
where the symbol $\Delta$ and the subscript 'av' represent the differences and averages of the tangential electric or magnetic fields at both sides of the metasurface, and $\overline{\overline{ \chi}}_\text{ee}$, $\overline{\overline{ \chi}}_\text{mm}$, $\overline{\overline{ \chi}}_\text{em}$, $\overline{\overline{ \chi}}_\text{me}$ are the bianisotropic susceptibility tensors characterizing the metasurface.

 Figure~\ref{fig:Concepts}(b) plots the FDFD-simulated~\cite{vahabzadeh2016simulation} fields propagating across a uniform nonreciprocal phase shifting metasurface with phase difference $\Delta \phi = 180^\circ$, corresponding to a spatial gyrator.  The corresponding susceptiblities for an $x$-polarized incident wave are obtained from~\eqref{eq:GSTC} as

\begin{equation}\label{eq:susceptibility_NR_phase_shifter}
\begin{split}
&\chi_\text{ee}^{xx} = \frac{\tan \left(\dfrac{\text{$\phi_1 $}}{2}\right)+\tan \left(\dfrac{\text{$\phi_2 $}}{2}\right)}{\eta  \omega  \epsilon }, \quad \chi_\text{mm}^{yy} = \dfrac{\eta  \left(\tan \left(\dfrac{\text{$\phi_1$}}{2}\right)+\tan \left(\dfrac{\text{$\phi_2$}}{2}\right)\right)}{\mu  \omega }, \\ &\chi_\text{em}^{xy} = \dfrac{\tan \left(\dfrac{\text{$\phi_1 $}}{2}\right)-\tan \left(\dfrac{\text{$\phi_2 $}}{2}\right)}{k}, \quad \chi_\text{me}^{yx} = \dfrac{\tan \left(\dfrac{\text{$\phi_1 $}}{2}\right)-\tan \left(\dfrac{\text{$\phi_2 $}}{2}\right)}{k},
\end{split}
\end{equation}
where $\phi_1$  and $\phi_2$ correspond to the phase shifts in the opposite directions of propagation.

Figure~\ref{fig:Concepts}(c) plots the full-wave simulated scattering parameters for a transistor-based implementation of a uniform nonreciprocal phase shifting metasurface. The unit cell of this metasurface is composed of three metallic layers supported by two Rogers3003 dielectric substrates and has a period of $\sim \lambda_0 /4.5$ at the operating frequency. A transistor (modeled by an ideal isolator) connects the first and third metallic layers with the second layer, which is acting as the ground. The unit cell provides a $90^\circ$ phase difference between the forward and backward transmissions with around $-3$~dB transmission. Optimization work is still required for higher performance.

\vspace{-2.5mm}
\section{Numerical Demonstration with a Spatial Circulator}

\begin{figure}
  \centering
  \includegraphics[width=0.8\textwidth]{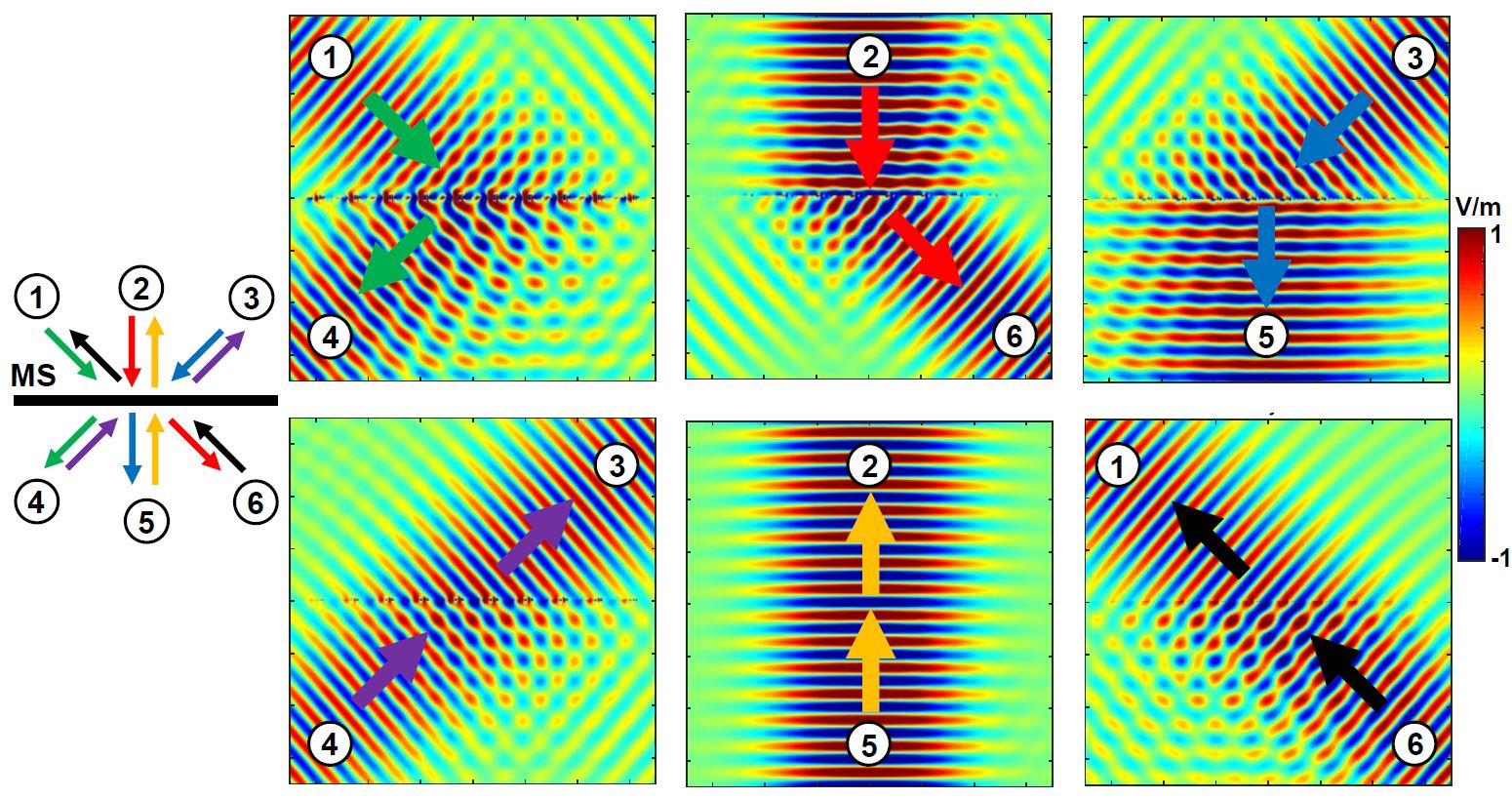}
  \caption{FDFD-simulated of real part of $E_x$ of a nonreciprocal phase gradient metasurface corresponding to the design of~\figref{fig:Concepts}~(a). This specific design corresponds to a spatial 6-port circulator, but many other devices are possible.}\label{fig:circulator}
\end{figure}

Figure~\ref{fig:circulator} presents FDFD simulation results for the nonreciprocal phase gradient metasurface of~\figref{fig:Concepts}(a) for a phase gradient period of $\sqrt{2} \lambda_0$ corresponding to the $45^\circ$ deflection of a normally incident wave. In this simulation setup, 6 ports are placed in the directions corresponding to the 6 propagating diffraction orders of the supercell period. This metasurface operates as a 6-port spatial circulator with circulation following the sequence $1 \rightarrow 4 \rightarrow 3 \rightarrow 5 \rightarrow 2 \rightarrow 6 \rightarrow 1$. This only represents an example; other devices can be realized by appropriately designing the forward and backward phase gradients, and even manipulating the transmission and reflection magnitudes.

Another promising application of the proposed nonreciprocal phase gradient metasurfaces is the realization of antennas with nonreciprocal radiation patterns~\cite{zang2015relay,hadad2016breaking} . For instance, an antenna placed at the port 5 in~\figref{fig:circulator} would transmit straightforwardly towards port 2, but receive from port 3 at $45^\circ$.

While the transistor-based implementation demonstrated in~\figref{fig:Concepts}(b) is currently restricted to microwave regime, it may soon extend to the optical regime following recent advances in the development of optical transistors~\cite{sun2018single}. In the meanwhile, the proposed nonreciprocal phase gradient concept can be readily applied at optical frequencies using alternative nonreciprocal technologies such as for instance space-time modulation~\cite{caloz2018electromagnetic}.

\vspace{-2.5mm}
\section{Conclusion}
\vspace{-2.5mm}
We have presented the concept of nonreciprocal phase gradient metasurfaces based on nonreciprocal phase shifting unit cells, proposed a related transistor-based implementation, and demonstrated the concept in a 6-port spatial circulator device. We have also pointed out that this technology has a potential for a diversity of other applications.
{\small

\bibliography{LIB}
\bibliographystyle{ieeetr}

}

\end{document}